\documentclass[twocolumn,showpacs,preprintnumbers,amsmath,amssymb]{revtex4}

\usepackage{graphicx}
\usepackage{dcolumn}
\usepackage{bm}

\begin{document}

\preprint{APS/123-QED}

\title{INVESTIGATION OF TRAVELING IONOSPHERIC DISTURBANCES DURING A MIDLATITUDE SPREAD F EVENT}
\author{G. Smith and A. Kebede}
\email{abkebede@gmail.com}

\affiliation{Department of Physics}
\address{North Carolina Agricultural and Technical State University}
\address{1601 E. Market St., Greensboro, NC 27411 USA}

\date{\today}
\begin{abstract}
Abstract: During a midlatitude spread F (MSF) event, data was
collected to investigate the circumstances that may lead to MSF.
Using the Total Electron Content (TEC) derived from the NCAT-SCINDA
GPS station and the Continuous Operating Reference Stations (CORS)
Traveling Ionospheric Disturbances (TID) were analyzed during a
period of MSF over Wallops Island, Virginia.  In addition to the TEC
analysis, scintillation calculations have been made using the
NCAT-SCINDA GPS receiver, USRP receiver and a Narrow Band (NB)
receiver.   Scintillation levels on the GPS, USRP and NB signals
were very low throughout the period of MSF.  Analysis of TEC data
from multiple CORS sites has shown the presence of medium scale
atmospheric gravity waves (AGW) within the MSF event region
propagating towards low latitudes with a small eastward component.
This is consistent with theories showing AGW may lead to MSF if an
oppositely directed neutral wind is present.  This study was
performed in conjunction with a sounding rocket experiment
investigating ionospheric disturbances at multiple scale sizes.

\end{abstract}


\maketitle
\section{INTRODUCTION}
The University of Texas (UT) performed a study using ionogram data
from Wallops Island, VA (WI). Their study concluded that irregular
variations at midlatitudes occur several times throughout the year
with increased frequency during the fall and winter seasons
\cite{REF1}.  As a follow-up to the UT's previous WI, study the UT
proposed an experiment where a sounding rocket would be launched
from WI into the F layer during a spread F event.  The sounding
rocket collected electric field, neutral wind, and ion density
measurements. In conjunction with this experiment NCAT collected
scintillation and total electron content data at GPS frequencies
from its campus located in Greensboro, North Carolina. Scintillation
data was also collected by the Air Force Research laboratory (AFRL)
at VHF from Martha's Vineyard (MV) (Figure 1). Midlatitude spread F
was confirmed by digisonde measurements (Figure 2) and the sounding
rocket was launched at 12:12 AM local time on October 30, 2007 from
Wallops Island, Virginia (WI).

\begin{figure}[htb]
\center{\includegraphics[scale=0.4]{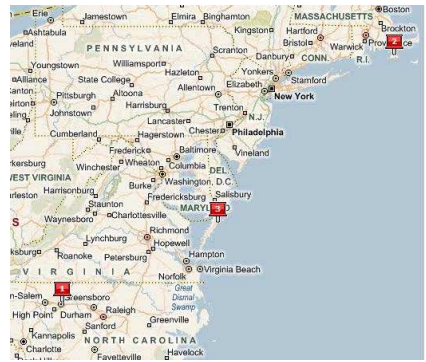}}
\caption{1-Greensboro,NC 2- Martha's Vineyard, MA 3- Wallops
Island,VA} \label{Fig}
\end{figure}

\begin{figure}[htb]
\center{\includegraphics[scale=0.4]{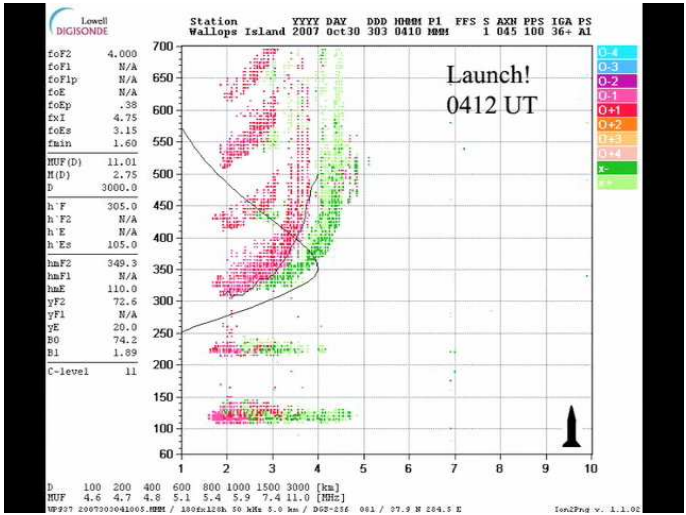}} \caption{Ionogram
showing spread F and sporadic E at Wallops Island, VA just before
launch} \label{Fig}
\end{figure}

\section{Methodology and data analysis}

GPS satellites are distributed throughout the sky giving each PRN a
unique line of site to the receiver.  This requires that PRNs be
selected based on the path that they traverse; this included
eliminating PRNs that were not within the WI area (space and time).
The PRNs that were close to the WI area around the launch were PRNs
2, 4, 5, 10, 12, 17 and 23 (Figure 3).

\begin{figure}[htb]
\center{\includegraphics[scale=0.45]{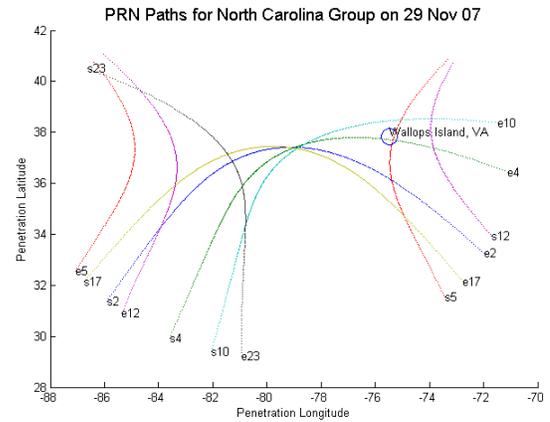}}
\center{\includegraphics[scale=0.4]{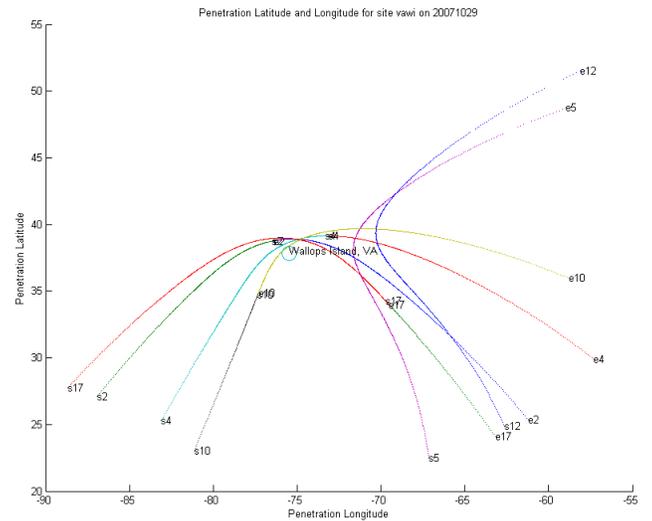}}
\caption{NorthCarolina and Virginia group paths for PRNs close to
Wallops Island, VA on 29 Oct 07} \label{Fig}
\end{figure}

The PRNs that were close to WI during the launch into the spread F
event were 2, 4 10, and 17, with PRN 2, 4, and 10 passing within one
degree latitude and longitude of WI.

\section{Scintillation Analysis Results}

The AFRL collected VHF scintillation data using a Universal Software
Radio Peripheral (USRP) receiver operating on four frequencies, and
a separate hardware based VHF receiver.  NCAT collected L band
scintillation data using its GPS based SCINDA receiver. VHF and L
band scintillation is expressed as the S4 index following the
relation; $S_4 =  \frac{\sqrt{\langle I\rangle ^2-\langle I \rangle
^2}}{\langle I \rangle}$

\begin{figure}[htb]
\center{\includegraphics[scale=0.4]{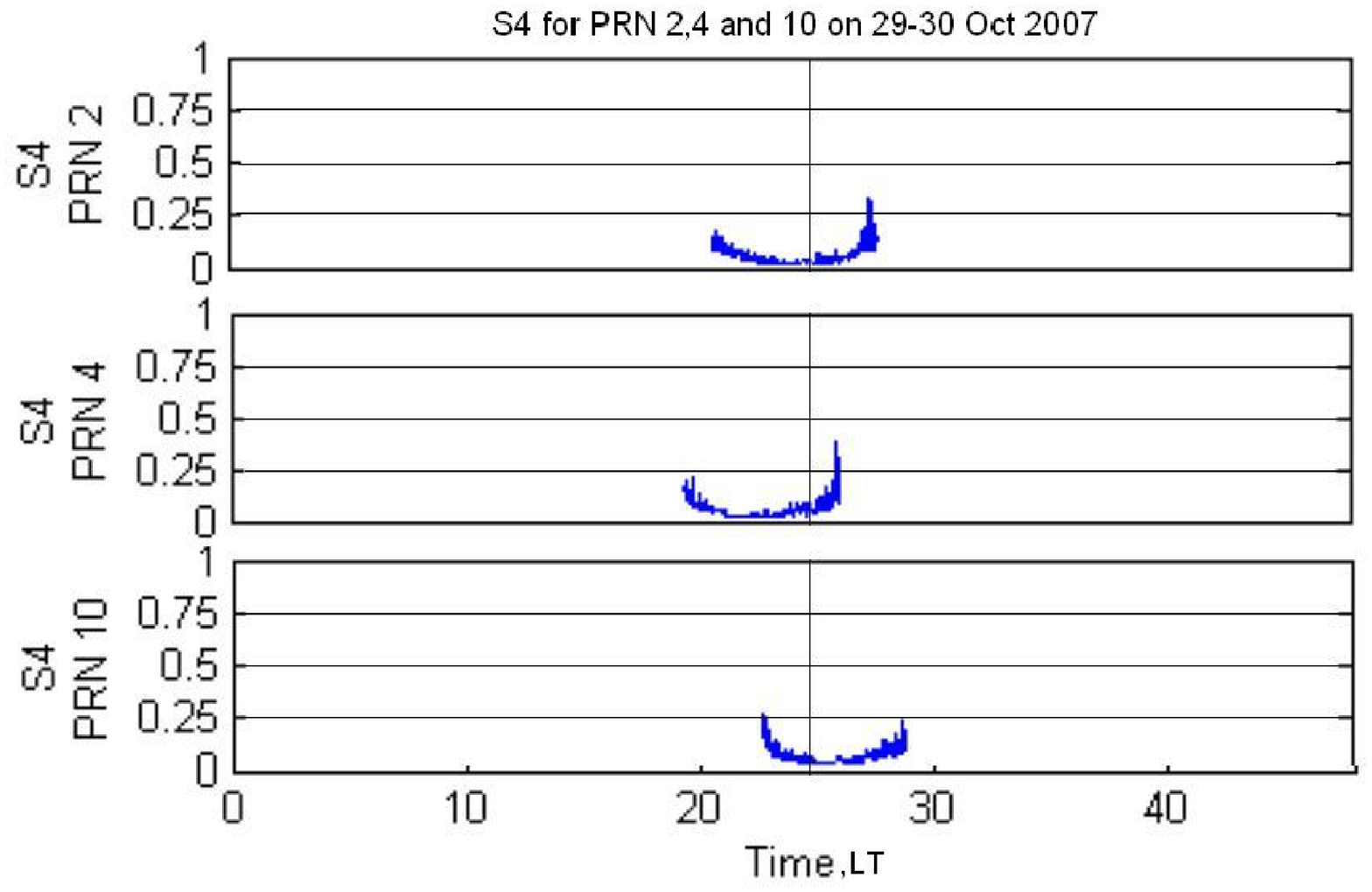}}
\caption{Scintillation index measurements from NCAT for PRNs 2, 4,
and 10 (Vertical line represents the end of 29 Oct 2007)}
\label{Fig}
\end{figure}

\begin{figure}[htb]
\center{\includegraphics[scale=0.5]{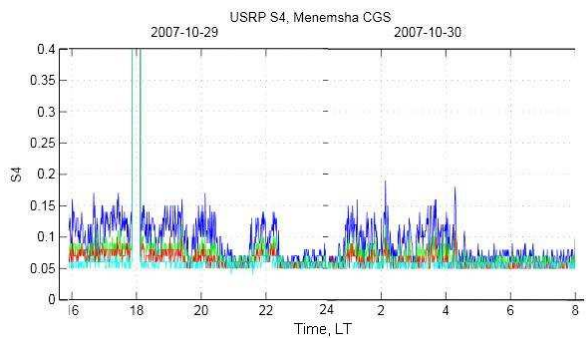}}
\caption{Scintillation index measured at four frequencies using the
USRP from Martha's Vineyard} \label{Fig}
\end{figure}

Scintillation was observed for many days before the rocket launch
from both NCAT and MV; S4 did not exceed 0.2 on either the USRP or
VHF receiver and did not exceed 0.4 for elevations higher than 30
degrees on the SCINDA receiver. During the launch, scintillation
observed at NCAT did not exceed 0.2 for PRNs 2, 4 and 10 during the
flight of the rocket experiment (Figure 4). Scintillation
measurements from MV during the rocket experiment were similar and
did not exceed 0.2 for a period of eight hours centered on the
launch time (Figure 5). The low levels of scintillation suggest that
small-scale ionospheric irregularities played a limited role during
this MSF event\cite{REF2}.

\section{Traveling Ionospheric Disturbance Analysis}
Total Electron Content (TEC) data was obtained from the SCINDA GPS
receiver and the CORS network.  Two groups of GPS receivers were
used to analyze ionospheric disturbances; The Virginia group (VG)
was centered on the rocket launch site at WI and the North Carolina
Group (NCG) was centered on the NCAT campus.

\section{TEC from the CORS Network}
The CORS network is organized by NOAA which makes data available
online in Receiver Independent Exchange (RINEX) format. This format
stores the pseudoranges and carrier phases for each PRN.  In order
to calculate TEC from the pseudoranges and carrier phases, the
azimuth and elevation of each satellite must be known. This was
calculated with software provided by the Low Latitude Ionospheric
Sensor Network (LISN) which made use of almanac data to calculate
the GPS constellation orbital path.  TEC was also calculated with
software provided by the LISN which required the azimuth, elevation
and either the DPR1-DCP1 or DPR1-DPR2 satellite bias information.

\section{Filtering TEC}
TEC data was filtered to remove ionospheric structures that were
larger than mesoscale objects.  This was accomplished by fitting a
small term sine function of the form $TEC_f=Asin(Bt+C)+Dsin(Et+F)+G$
and subtracting. $A$, $B$, $C$, $D$, $E$, $F$, and $G$ are
constants. The small term sine function allowed for the removal of
diurnal and tidal variations and geometrical aberrations. This
method was highly successful when using TEC data that was calibrated
using different methods (SCINDA and LISN). Figure 6 shows that this
method allows mesoscale ionospheric structures to be compared
between data sets that would otherwise not be discernable.

\begin{figure}[htb]
\center{\includegraphics[scale=0.25]{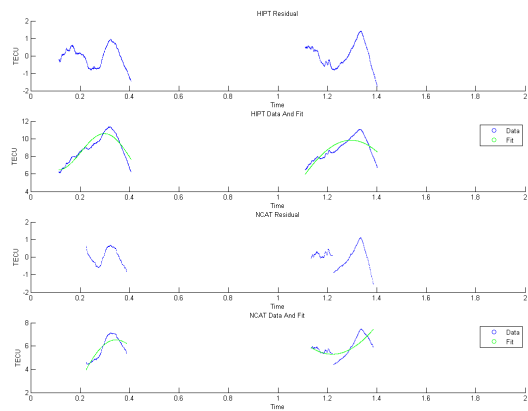}} \caption{Comparison
of NCAT SCINDA Calibrated TEC and HIPT  LISN Calibrated TEC}
\label{Fig}
\end{figure}

\section{Results of TID Analysis}

TEC was filtered on PRNs where data was collected during the rocket
launch; these were PRNs 2, 4, 5, 10, 12, and 17. The NCG PRNs that
were within two degrees latitude and longitude (LL) during the
launch were PRN 2, 4, 10, and 17 with PRN 4 passing directly over
the launch site. The VG PRNs that were within two degrees LL during
the launch were 2, 4, 10, and 17 with PRN 10 passing directly over
the launch site. As indicated in Figure 7, the TEC analysis showed
the presence of an atmospheric gravity wave (AGW) with a period on
the order of one and one half hours with peek to peek amplitude of
three TECU occurring just before midnight on 29 October 2007.
\begin{figure}[htb]
\center{\includegraphics[scale=0.4]{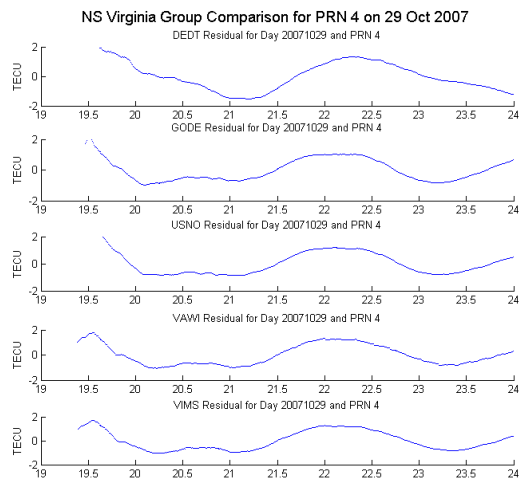}} \caption{North to
South VG Comparison for PRN 2} \label{Fig}
\end{figure}

VG and NCG comparison showed that the AGW was southward propagating
with a small eastward component (Figure 8).

\begin{figure}[htb]
\center{\includegraphics[scale=0.4]{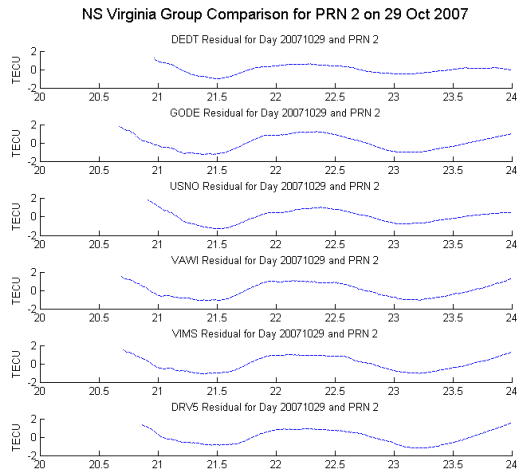}} \caption{East to West
VG Comparison for PRN 2} \label{fig:Fig}
\end{figure}

In Figure 2, sporadic E can be seen on the ionogram during the
launch at approximately 120 Km. Sporadic E has been shown to be
associated with AGW with periods from 10 to 70 minutes \cite{REF3}.
While the observed AGW has a period longer than 70 minutes;
numerical studies have shown that spatial resonance is not required
to modulate the height of sporadic E \cite{REF4}.

\section{Conclusion and Discussion}

We have shown scintillation levels were low throughout the MSF event
both at VHF and l band frequencies, suggesting that small-scale
ionospheric structures were minimal for the period of observation.
Also presented is a method to compare mesoscale ionospheric
structures between GPS-derived TEC data sets that have been
calibrated using different methods.  We have shown the presence of
an AGW with a period of approximately 90 minutes and amplitude of
three TECU. Furthermore the AGW was seen to propagate towards low
latitudes with as small eastward component. This small sine term
method should be expanded to other MSF events to calculate
correlation values between this method's ability to show mesoscale
ionospheric structures in relation to MSF and sporadic E.  The
selection of CORS sites should be expanded to give better
estimations of AGW velocity.

\begin{acknowledgments}
 This work was supported by the NASA MUCERPI2003 under the Grant NNG04GD63G.
 We are grateful to Dr. Keith Groves and the Air Force Research Laboratory where Mr. Galen Smith was
 employed as a summer graduate student to complete his Masters Thesis  work.

\end{acknowledgments}

\end{document}